\documentclass[preprint,nofootinbib]{revtex4}

\usepackage{graphicx}

\begin{document}

\title{On vacuum density, the initial singularity and dark
energy\footnote{Second Award in the 2009
Essay Competition of the Gravity Research Foundation.}}

\author{Saulo Carneiro$^{1,2}$\footnote{ICTP Associate Member. E-mail address: saulo.carneiro@pq.cnpq.br.} and
Reza Tavakol$^{1}$\footnote{E-mail address: r.tavakol@qmul.ac.uk.}}

\affiliation{$^1$Astronomy Unit, School of Mathematical Sciences,
Queen Mary University of London, Mile End Road, London E1 4NS,
UK\\$^2$Instituto de F\'{\i}sica, Universidade Federal da Bahia,
Salvador, BA, 40210-340, Brazil}

\begin{abstract}
Standard cosmology poses a number of important questions. Apart from
its singular origin, it possesses early and late accelerating phases
required to account for observations. The vacuum energy has been
considered as a possible way to resolve some of these questions. The
vacuum energy density induced by free fields in an early de Sitter phase
has earlier been estimated to be proportional to $H^4$, while more recently
it has been suggested that the QCD condensate induces a term proportional to $H$ at late
times. These results have been employed in models which are
non-singular and inflationary at early times and accelerating at
late times. Here we cast these models
in terms of scalar fields and study the corresponding
spectrum of primordial perturbations.
At early times the spectrum is found to be not scale-invariant,
thus implying that slow-roll inflation is still required after the
phase transition induced by the vacuum.
At late times the corresponding scalar-field potential is
harmonic, with a mass of the order of the Hubble scale, a result
that may be understood in the light of the holographic conjecture.
\end{abstract}

\maketitle

Recent years have witnessed a tremendous accumulation of high
resolution data in cosmology. This has led to the so called standard
model, which poses a number of important questions. Apart from
having a singular origin, at which laws of physics break down, it
also possesses accelerating phases at early and late times. This has
resulted in numerous attempts at generalising the gravitational and
energy sectors of GR in order to account for these features.
Interestingly all viable models have proved to be very close to the
$\Lambda$CDM.

An important ingredient expected from
quantum considerations is the vacuum energy,
which has often been thought to be responsible for the
origin of the cosmological parameter $\Lambda$.
As is well known, it is not easy to obtain the vacuum density in curved
spacetimes, even in the simple case of a scalar field \cite{Ford,Dowker,Davies}. An exception
is the case of fields with conformal
invariance in de Sitter spacetime \cite{Dowker,Davies}, for which the renormalized vacuum
density has been found to be proportional to $H^4$, where $H$ is the Hubble parameter.
In a general FLRW background, some
ambiguities can be fixed by imposing the conservation of
the vacuum energy-momentum tensor \cite{Davies}.
This allows a general expression for the vacuum density
to be derived, which was in fact originally used by
Starobinsky in his pioneering inflationary model \cite{Starobinsky}.

Recently the de Sitter ansatz $\Lambda \propto H^4$ was employed in a
quasi-de Sitter setting in order
to construct a non-singular scenario, with a phase transition
between a past-eternal de Sitter phase and
a radiation-dominated epoch \cite{Carneiro}.
An underlying assumption was that a dynamical vacuum
interacts with matter, since only the
conservation of the total energy-momentum tensor is implied by
Einstein equations. Thus in this model the decaying vacuum
acts as a source of relativistic matter during the expansion.

On the other hand, it has been suggested that
the QCD condensate induces a vacuum density
proportional to $H$ at late times \cite{Schutzhold,Klinkhamer}.
This has also been employed to construct models
to account for the late acceleration of the universe.

In this paper we formulate these ansatz
in terms of self-interacting scalar fields in a spatially flat FLRW
spacetime\footnote{This procedure can similarly be applied to
other ansatz.}.
In this framework the Lagrangian takes the usual form
\begin{equation} \label{Lagrangian}
{\cal L} = \sqrt{-g}\left[\frac{R}{2}-\frac{1}{2}(\partial \phi)^2-V(\phi)\right],
\end{equation}
with the corresponding field equations given by
\begin{eqnarray}
3 H^2 &=& V + 2H'^2, \label{field equation 1}\\
\dot{\phi}&=&-2H'. \label{field equation 2}
\end{eqnarray}
Here the prime and the dot denote derivatives with respect to
$\phi$ and the cosmological time $t$, respectively.

We begin with the early phase and consider a potential
proportional to $H^4$,
which has a maximum in the asymptotic de Sitter limit.
The energy density and pressure of this field are
given by $\rho = V + \dot{\phi}^2/2$ and
$p = -V + \dot{\phi}^2/2$, respectively.
Therefore, we can interpret the material content as formed by a
vacuum term with density $V$ and pressure $-V$, plus a
stiff fluid with density and pressure given by $\dot{\phi}^2/2$. As the
transition from the de Sitter phase takes place, with the field
rolling down the potential, the energy stored in the vacuum term is
transferred to the stiff component, and eventually converted into
matter fields by a suitable coupling, as in the usual
reheating mechanism.

\begin{figure}[t]
\begin{center}
\includegraphics[height=6cm,width=10cm]{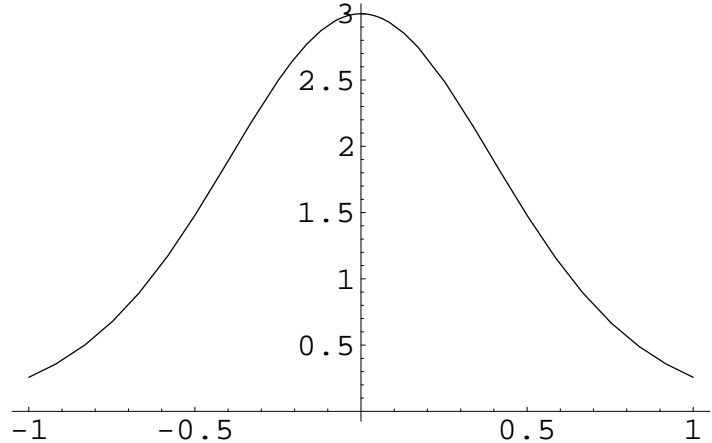}
\end{center}
\caption{The scalar field potential corresponding to Eq.~(\ref{ansatz})
as a function of $\phi$ ($\sigma= 1$).}
\end{figure}

With the ansatz
\begin{equation} \label{ansatz}
V= 3 \sigma H^4,
\end{equation}
where $\sigma$ is a positive constant, it is straightforward,
using Eqs. (\ref{field equation 1})-(\ref{field equation 2}), to find the solution
\begin{equation} \label{early Hubble}
H = \frac{2\,e^{\sqrt{3/2}\phi}}{1+\sigma e^{\sqrt{6}\phi}}.
\end{equation}
The corresponding potential, with $\sigma=1$, is shown in Figure 1.

Also, using $\dot{H}=H'\dot{\phi}$ in Eqs. (\ref{field equation 1})-(\ref{field equation 2}),
we obtain an evolution equation for $H$,
\begin{equation} \label{evolution equation}
\dot{H}+3H^2-3\sigma H^4=0.
\end{equation}
This equation has the equilibrium point $\sigma H^2 = 1$, which
corresponds to a de Sitter universe with
energy scale $\sigma^{-1/2}$ in Planck units. This solution
is, however, unstable, as indicated by the solution
\begin{equation} \label{early Hubble x time}
\tilde{t}=\frac{1}{\tilde{H}}-\tanh^{-1}\tilde{H},
\end{equation}
where we have introduced the re-scaled quantities
$\tilde{H}=\sqrt{\sigma}H$ and $\tilde{t}=3t/\sqrt{\sigma}$, and
have conveniently chosen the integration constant.

\begin{figure}[t]
\begin{center}
\includegraphics[height=6cm,width=10cm]{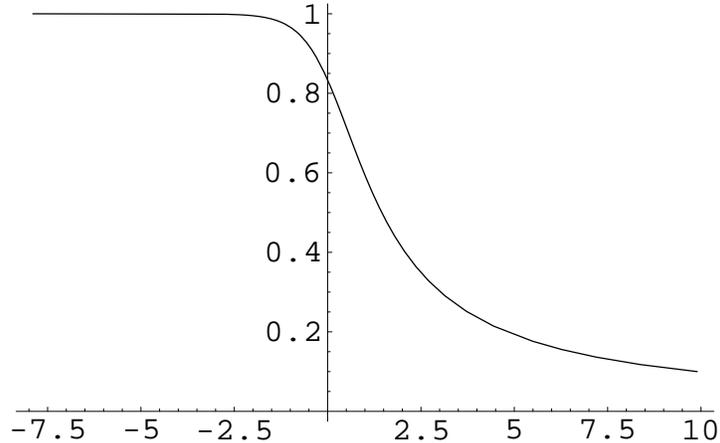}
\end{center}
\caption{The Hubble parameter as a function of time ($\sigma = 1$).}
\end{figure}

The time evolution of $\tilde{H}$ is depicted in Figure 2. As $t
\rightarrow - \infty$ the solution approaches
the de Sitter point, with $\sigma H^2 = 1$, as expected.
The solution remains quasi-de Sitter during an indefinitely long
time. Then, around $t = 0$, it goes through a phase transition, with $H$
and $V$ decaying very fast. It is worth noting that for
$\sigma > 1$ we have $H < 1$ during the entire evolution, thus we never
enter a trans-Planckian regime, where the semi-classical, one-loop
derivation of the vacuum density would not be valid.

This model has a number of appealing features. In addition
to being non-singular, the presence of a quasi-de Sitter phase
solves some of the problems
usually addressed by inflation, such as the horizon and flatness problems.
However, the most important outcome of an inflationary phase
is to produce a scale invariant spectrum of primordial perturbations.
To be viable as an inflationary scenario, it is therefore
important to check whether
the above model can also produce such a scale-invariant spectrum.

The evolution equation for perturbations in the scalar field is given by \cite{Dodelson}
\begin{equation} \label{perturbations}
\delta {\phi}^{**} + 2aH\delta {\phi}^* + (k^2 + a^2 V'') \delta\phi = 0,
\end{equation}
where $a$ is the scale factor, $k$ is the co-moving wave number of a given mode,
and * denotes a derivative with respect to the
conformal time $\eta$.

From (\ref{ansatz}), (\ref{early Hubble}) and (\ref{field equation 2}) it is possible to obtain
\begin{eqnarray}
V'' &=& 18\sigma H^4 (4 - 5\sigma H^2) \label{potential},\\
\epsilon &=& 3(1-\sigma H^2) \label{epsilon},
\end{eqnarray}
with $\epsilon = d(H^{-1})/dt$. In the quasi-de Sitter phase we have
$\sigma H^2 \approx 1$ and, hence, $V'' \approx -18 H^2$ and
$\epsilon \ll 1$. By using $Ha \approx -1/\eta$ and introducing
$\delta \tilde{\phi} = a \delta \phi$, the perturbation
Eq.~(\ref{perturbations}) can be rewritten as
\begin{equation} \label{perturbations 2}
\delta{\tilde{\phi}}^{**} + \left(k^2 -\frac{20}{\eta^2}+ \frac{30\epsilon}{\eta^2}\right) \delta\tilde{\phi} = 0.
\end{equation}
Neglecting the term in $\epsilon$, the appropriately normalized solution has the form
\begin{equation} \label{delta phi}
\delta\tilde{\phi} \approx \frac{\sqrt{-\pi \eta}}{2}\,H_{\frac{9}{2}}^{(1)}[-k\eta],
\end{equation}
where $H_n^{(1)}$ is the Hankel function of the first kind. For $k \rightarrow \infty$,
this solution reduces to $e^{-ik\eta}/\sqrt{2k}$, as expected.
On the other hand, for $k\eta=-1$, when the mode crosses the horizon,
we have $|\delta \tilde{\phi}| \approx 10^2/\sqrt{2k}$.

The power spectrum of scalar perturbations in the metric is given by \cite{Dodelson}
\begin{equation} \label{power spectrum 1}
P_{\Psi} = \frac{4}{9}\left(\frac{aH}{{\phi}^*}\right)^2 \left| \delta \phi \right|^2_{k\eta=-1}.
\end{equation}
Using (\ref{delta phi}) and $(aH/{\phi}^*)^2 = (2\epsilon)^{-1}$, we have
\begin{equation} \label{power spectrum 2}
k^3P_{\Psi} \approx \frac{10^4\, H^2}{9\epsilon},
\end{equation}
with the right hand side evaluated at the horizon crossing. Apart from the
factor of $10^4$, this is the same expression we find in slow-roll
inflation. The power spectrum of tensor modes is the same, and hence
the extra factor of $10^4$ leads to a stronger suppression of
primordial gravitational waves. Nevertheless, $\epsilon$ here has a
strong dependence on $k$, which makes the spectrum scale-dependent.
Indeed, from (\ref{epsilon}) it is possible to show that, for
$\sigma H^2 \approx 1$, $\epsilon \propto \eta^{-6}$. Therefore, at
the horizon crossing we have $\epsilon \propto k^6$, leading to a
scalar spectral index $n - 1\approx -6$. This result follows
from the fact
that we do not have a slow-roll potential, since from (\ref{ansatz})
and (\ref{potential}) we have, for $\sigma H^2 \approx 1$, $V''/V
\approx -6$. Therefore, this scenario cannot by itself produce a
scale-invariant spectrum. To achieve this, the phase transition
described here must be followed by a usual inflationary epoch.

Let us now discuss the potential role of the vacuum energy at late
times. The free-field vacuum density, of order $H^4$, is presently
too small to be identified with the dark energy. On the other hand,
any contribution proportional to $H^2$ may be absorbed by the
gravitational constant in Einstein equations \cite{Parker}. It is
usually believed that, apart from terms in $H^2$ and $H^4$, the only
remaining contribution is a free constant, which may be absorbed by
a cosmological constant. Nevertheless, it has recently been claimed
that the introduction of interactions may change this picture
\cite{Schutzhold,Klinkhamer}. In particular, the QCD condensate
leads to a vacuum density of order $m^3 H$, where $m \approx 150$
MeV is the energy scale of the QCD phase transition\footnote{A term
linear in $H$ also appears in models with a modified Friedmann
equation, in the context of high-dimensional theories. See, for
example, reference \cite{Turner}.}. Although not conclusive, this
result gives the correct order of magnitude for the cosmological
term, $\Lambda \approx m^6$, and, as a byproduct, the Dirac large
number coincidence $H \approx m^3$.

At present it is not known whether this linear relation
is valid only in the
de Sitter limit or can also hold in the dynamical regime near this limit.
Assuming the latter, again
in order to ensure the conservation of total energy,
matter production would be required. This scenario was studied recently,
by confronting its predictions with the current observations.
A joint analysis of SNe Ia, baryonic acoustic oscillations and the
position of the first peak of CMB anisotropy spectrum produced a good
concordance \cite{Jailson}. However, the production of dark matter leads
to a power suppression in the mass spectrum, which seems to rule out this
scenario \cite{Julio}. This difficulty may be overcome if we avoid matter production by associating
the varying $\Lambda$ with a quintessence field, as was done in the
early-time limit above. A full study, which also includes matter,
is in progress. Here we shall consider the simpler case where only the
quintessence field is present, which is reasonable very near the de
Sitter limit.

With the ansatz $V = 3\sigma H$, where now $\sigma \approx m^3$,
Eqs.~(\ref{field equation 1})-(\ref{field equation 2}) have the
solution
\begin{equation} \label{late Hubble}
V = \frac{3\sigma\left(e^{\sqrt{3/2}\phi}+\sigma\right)^2}{4e^{\sqrt{3/2}\phi}},
\end{equation}
\begin{equation} \label{late Hubbe x time}
H = \frac{\sigma e^{3\sigma t}}{e^{3\sigma t}-1}.
\end{equation}
As can be seen, as $t \rightarrow \infty$, $H \rightarrow \sigma$
as expected.

The potential possesses a minimum at $\phi_0 = 2\ln\sigma/\sqrt{6}$,
corresponding to the de Sitter limit. Around this point the potential
can be expanded as
\begin{equation} \label{potential expansion}
V \approx V_0 + \frac{M^2}{2} \left (\phi - \phi_0 \right ) ^2,
\end{equation}
where $V_0 = 3\sigma^2$ and $M=3\sigma/2$. The mass term is worthy of note.
In the de Sitter limit we have $\sigma = H = \sqrt{\Lambda/3}$. Therefore, the mass of the quintessence field is given by
\begin{equation} \label{mass}
M = \frac{3}{2}\sqrt{\Lambda/3}.
\end{equation}
This is the mass expected for elementary degrees of freedom in the
context of the holographic conjecture \cite{Marugan}.
Indeed, recalling the observable mass in the universe, $E \approx
\rho/H^3 \approx 1/\sqrt{\Lambda}$, and the entropy associated with
the de Sitter horizon, $N \approx 1/H^2 \approx 1/\Lambda$, we
obtain $M \approx E/N \approx \sqrt{\Lambda}$ for the mass of each
degree of freedom.

In conclusion, we note that the vacuum energy, usually considered a
problem in quantum field theories and cosmology, may actually shed
new light on other fundamental problems, such as the initial
singularity and the nature of dark energy. At early times the
free-field contribution to the vacuum density, proportional to
$H^4$, leads to a non-singular scenario, which in this paper was
modeled by a self-interacting scalar field, with the potential
playing the role of the vacuum term. In this context, it was
possible to show that the primordial transition from the quasi-de
Sitter phase does not produce a scale-invariant spectrum of
perturbations, for which a subsequent, usual inflationary epoch is
required. On the other hand, at late times we have associated the
vacuum term with a quintessence field with potential linear in $H$,
as suggested by QCD results in the de Sitter space-time. The
presence of such a field allows the conservation of the total energy
without invoking matter production. Near the future de Sitter limit,
the quintessence potential is shown to be harmonic, with a mass term
in agreement with the holographic prescription.

\section*{Acknowledgements}
We wish to thank James Lidsey and Raul Abramo for very
helpful discussions. Saulo Carneiro was partially supported by CAPES
(Brazil).

%\newpage


\begin{thebibliography}{}

\bibitem{Ford} L. H. Ford, Phys. Rev. {\bf D 11}, 3370, 1975.

\bibitem{Dowker} J. S. Dowker and R. Critchley, Phys. Rev. {\bf D 13}, 3224, 1976.

\bibitem{Davies} P. C. W. Davies, Phys. Lett. {\bf B 68}, 402, 1977.

\bibitem{Starobinsky} A. A. Starobinsky, Phys. Lett. {\bf B 91}, 99, 1980.

\bibitem{Carneiro} S. Carneiro, Int. J. Mod. Phys. {\bf D 15}, 2241, 2006.

\bibitem{Schutzhold} R. Schutzhold, Phys. Rev. Lett. {\bf 89}, 081302, 2002.

\bibitem{Klinkhamer} F. R. Klinkhamer and G. E. Volovik, Phys. Rev. {\bf D 79}, 063527, 2009.

\bibitem{Dodelson} S. Dodelson, Modern Cosmology (Academic Press, 2003).

\bibitem{Parker} S. A. Fulling and L. Parker, Phys. Rev. {\bf D 10}, 3905, 1974.

\bibitem{Turner} G. Dvali and M. Turner, astro-ph/0301510.

\bibitem{Jailson} S. Carneiro, M. A. Dantas, C. Pigozzo, and J. S. Alcaniz, Phys. Rev. {\bf D 77}, 083504, 2008.

\bibitem{Julio} H. A. Borges, S. Carneiro, J. C. Fabris, and C. Pigozzo, Phys. Rev. {\bf D 77}, 043513, 2008.

\bibitem{Marugan} G. A. Mena Marug\'an and S. Carneiro, Phys. Rev. {\bf D 65}, 087303,
2002; R. Tavakol and G. F. R. Ellis, Phys. Lett. {\bf B 469}, 37, 1999; R. Bousso, Rev. Mod. Phys. {\bf 74}, 825, 2002.

\end{thebibliography}
\end{document}